\documentstyle[aps,prl,multicol,epsf]{revtex}

\def\8{\infty}

\def\undertext#1{\vtop{\hbox{#1}\kern 1pt \hrule}}

\def\VEV#1{\left\langle\,#1\,\right\rangle}

\def\br{\\ \nonumber & &}

\def\be{\begin{equation}}
\def\ee{\end{equation}}
\def\bea{\begin{eqnarray} & &}
\def\eea{\end{eqnarray}}
\def\ct#1{\cite{#1}}
\def\rf#1{(\ref{#1})}

\begin{document}
\draft

\title{Orthogonality Catastrophe and Spontaneous Symmetry Breaking
in Double-layer Fermi-liquid-like States}

\author {Victor Gurarie$^a$ and Yong Baek Kim$^b$}

\address{$^a$Institute for Theoretical Physics, University of California,
Santa Barbara, CA 93106-4030}
\address{$^b$Department of Physics, The Ohio State University, 
Columbus, OH 43210}

\date{\today}
\maketitle


\begin{abstract}

The double-layer electron system with total filling factor 
$\nu=1/2$ can be regarded as two separate Fermi-liquid-like 
states with $\nu=1/4$ when the layer separation is sufficiently large
and there is no tunneling. The weak tunneling in this state
suffers an orthogonality catastrophe and it becomes irrelevant.
Using the symmetric and antisymmetric combinations of 
layer indices as the pseudospin degrees of freedom, we show that
there exists the first order transition from the above
pseudospin unpolarized state to the pseudospin polarized
Fermi-liquid-like state with $\nu=1/2$ as the tunneling strength
becomes sufficiently large.

\end{abstract}


\begin{multicols}{2}

The compressible Fermi-liquid-like state at the filling factor
$\nu=1/2m$ in the lowest Landau level has been a subject 
of intensive theoretical \cite{hal} and 
experimental \cite{stormer} studies. 
Upon the experimental observation of
the Fermi surface effects of underlying quasiparticles,
the Chern-Simons Fermi-liquid theory of composite fermions was 
formulated and has been successful in explaining many 
qualitative features of experimental data \cite{HLR}.
The composite fermion consists of an electron
and even number, $2m$, of fictitious flux quanta
representing correlation holes around the electron \cite{jain}. 
The Chern-Simons Fermi-liquid theory of composite fermions 
is based on the 
idea that the averaged fictitious flux quanta cancels
the external magnetic field leading to the fermion
system in zero effective magnetic field.
Recently the truly lowest Landau level theory of the
composite fermions was developed and revealed
the dipolar nature of the 
quasiparticles \cite{shankar,dhlee,pasquier,read,stern}. 
It has been also established that the results in 
the long wavelength and low energy limit of 
this theory is equivalent to those of the 
original Chern-Simons Fermi-liquid 
theory \cite{shankar,dhlee,pasquier,read,stern}.     

On the other hand, the half-filled first 
Landau level (the total filling factor is $5/2$) 
supports a quantum Hall state \cite{willet}.
This observation leads to the studies of paired quantum Hall 
states which can be interpreted as the paired 
states of composite fermions \cite{HR,MR,greiter}. 
In particular, the spin polarized
Pfaffian Moore-Read (MR) state has arisen as a viable 
candidate which can be regarded as the $p$-wave
superconductor of composite fermions \cite{MR,greiter,RG}.
The mechanism for the pairing
was attributed to the modification of the short range
part of the electron-electron interaction in the
first Landau level \cite{rezayi}.

The paired quantum Hall states also arise in
the double-layer system of total filling factor 
$1/2m$. If two layers are well separated, the system 
consists of two separate $\nu=1/4m$ compressible 
Fermi-liquid-like states. 
It is found that when two layers become sufficiently close
to each other, there appears a pairing instability 
of composite fermions which belong to different 
layers \cite{bone}.
In particular, so-called (331) state may arise from $p$-wave
pairing of composite fermions with the total filling 
factor $\nu=1/2$ \cite{RG,halperin,ho,wen2}. 
If the symmetric and antisymmetric 
combinations of layer indices are used as the pseudospin 
degrees of freedom, the (331) state can be interpreted as
pseudospin unpolarized $p$-wave paired state of 
composite fermions \cite{RG,ho,wen2}. 
The tunneling in the (331) state
acts like a Zeeman magnetic field in the pseudospin space.  
Upon this observation, it was suggested that there 
exists a continuous transition to the pseudospin 
polarized MR state \cite{RG,ho,wen2}. 
In this case the transition
can be characterized by the appearance of finite 
pseudospin magnetization. 
   
Notice that the single layer spin polarized MR state 
in the first Landau level may occur due to the 
pairing instability of spin polarized Fermi-liquid
state of composite fermions \cite{rezayi}. 
Thus it is natural to ask
whether there exists the pseudospin polarized 
Fermi-liquid state of composite fermions in
the double layer system in the presence of
tunneling when the interlayer interaction
is not strong enough to derive the system to
a paired state.    

Let the layer indices $\uparrow$ and $\downarrow$
represent the eigenstates of the $z$-component 
of the pseudospin. 
If what tunnel between layers are {\it composite fermions}, 
the tunneling simply corresponds to applying a 
Zeeman magnetic field along the $x$-direction 
in the pseudospin space.
Thus it splits the energy of $\psi_+$ and $\psi_-$ 
fermions; $\psi_{\pm} = (\psi_{\uparrow} \pm
\psi_{\downarrow})/\sqrt{2}$ are
eigenstates of the $x$-component of the pseudospin. 
Any finite tunneling amplitude will give rise to pseudospin 
partially polarized states. When the tunneling amplitude 
becomes sufficiently large only the $\psi_-$ fermion states 
will be occupied leading to completely pseudospin polarized 
Fermi-liquid-like state with $\nu=1/2$. 

However, what tunnel between layers are {\it electrons}
and the above picture of the composite fermion tunneling 
may not be adequate.
Since the tunneling electrons should be represented as 
the composite fermions with four flux quanta, 
the tunneling also removes and adds flux quanta in each 
layer in addition to composite fermions. 
In the effective field theory description,
each tunneling event creates a monopole in the 
$a_- = a_{\uparrow} - a_{\downarrow}$ 
Chern-Simons gauge field configuration, where 
$a_{\uparrow}$ and $a_{\downarrow}$ represent 
the Chern-Simons gauge fields associated with 
the flux attachment transformations in 
different layers\cite{WZ,KW}. 
In this paper, we investigate the consequence of
the tunneling in the double-layer Fermi-liquid-like
states. In particular, we study the effect 
of monopoles on the tunneling between the layers and the 
transition from the pseudospin unpolarized state to 
the polarized Fermi-liquid-like state.

As previously noticed \cite{KW,he}, 
the simple tunneling process is 
suppressed due to the fact that the energy required to 
create a single monopole diverges (monopoles are confined).
This is an example of {\it orthogonality catastrophe}.
Thus the effect linear in the tunneling strength, $t$,
such as the energy splitting of $\psi_+$ and $\psi_-$ 
fermions, is absent.
The monopoles always appear in the monopole-antimonopole
pairs, with the probability proportional to 
$\exp(-S_{M \bar M})$ where $S_{M \bar M}$ is the 
monopole-antimonopole action which depends on the distance
between them in space-time.
We found that the effect of monopole-antimonopole 
pairs can be represented 
by an effective four fermion interaction. 
By studying this interaction,
we found that small tunneling amplitude, $t$, 
does not drive the transition to the polarized 
state due to the suppression of the tunneling events. 
On the other hand, when $t$ becomes sufficiently large, 
the effective Zeeman magnetic field is dynamically
generated and it leads to the polarized state.
Thus we conclude that, the transition between 
the pseudospin unpolarized state and polarized 
Fermi-liquid-like state is likely to be 
a first order transition. 

Let us begin with the discussion about composite fermions in 
separate compressible states in each layer. 
Here the filling factor of each layer is given by 1/4. 
In order to get compressible states, we attach four flux 
quanta to electrons in each layer independently.
This can be done by introducing two Chern-Simons gauge fields, 
$a_\uparrow$ and $a_\downarrow$. 
The action for the system can be written as 
(we set $\hbar=c=1$) \cite{HLR,bone}
\begin{eqnarray} 
\label{Lagrange}
S &=& S_0 + S_{\rm int} \ , \cr 
S_0 &=& \int  d^3 x \     
\psi^\dagger_{\uparrow} \left[  \partial_{\tau} +
a_{0 \uparrow}+ 
{\left( -i{\bf \nabla} + {\bf a}_{\uparrow} + 
{\bf A}\right)^2 \over 2 m^*} - \mu \right]
\psi_{\uparrow} \cr
&& + \ S_{CS}(a_{\uparrow}) \ + \  
\{ \uparrow \ \leftrightarrow \ \downarrow \} \ ,
\end{eqnarray}
where $S_{CS}$ is the Chern-Simons term,
\be
\label{Chern}
S_{CS}(a)={i \over 2 \pi \phi_0}
\int d^3 x \ \epsilon_{\alpha \beta \gamma} a_{\alpha}
\partial_{\beta} a_\gamma \ ,
\ee
whose role is to attach $\phi_0$ flux quanta to each 
composite fermion,
$S_{\rm int}$ denotes the interaction between electrons, 
$\psi$, $\psi^\dagger$ are the operators for the
{\it composite fermions}, $\mu$ is the chemical potential 
and ${\bf A}$ represents the external
magnetic field. If $\phi_0=4$, the external magnetic field is 
canceled by the mean value of 
${\bf a}_{\uparrow}$ and ${\bf a}_{\downarrow}$. From 
now on, we will use ${\bf a}_{\uparrow,\downarrow}$
to represent the fluctuations 
$\delta {\bf a}_{\uparrow,\downarrow} = 
{\bf a}_{\uparrow,\downarrow} - {\bf A}$.  

It is convenient to introduce 
${\bf a}_{\pm} = {\bf a}_{\uparrow}
\pm {\bf a}_{\downarrow}$ which are related to the 
symmetric and antisymmetric fluctuations of 
the electron densities as
$\nabla \times {\bf a}_{\pm} = 2 \pi \phi_0 
(\delta \rho_{\uparrow} \pm \delta \rho_{\downarrow})$. 
Now let us discuss what happens if we allow the tunneling between 
the layers. 
The tunneling term is given by
\be
S_{\rm t}= t \int d^3x \ \left [
\Psi^\dagger_\uparrow \Psi_\downarrow + 
\Psi^\dagger_\downarrow \Psi_\uparrow \right ] \ , 
\ee
where $\Psi^\dagger$, in contrast to $\psi^\dagger$, is the electron 
creation operator.
The fact that the electron consists of composite fermion
and four flux quanta can be represented by the 
following transformation,
\be
\label{flux}
\Psi^\dagger_\uparrow (x)
= e^{ i \int d^2 x' \ \varphi(x-x') \epsilon_{i j} \partial_i
a_{j \uparrow} (x') } \psi^\dagger_\uparrow (x) \ ,
\ee
where $i,j$ are the spatial indices, and
$\varphi(x)$ gives the angle between the two dimensional vector 
{\bf x} with respect to the $x$-axis.
Introducing $U$ for the flux creation
operator in the gauge field $a_-$, 
\be
\label{monopole}
U^\dagger(x) = e^{ i \int d^2 x' \ \varphi(x-x') 
\epsilon_{i j} \partial_i
a_{j -} (x') } \ ,
\ee
the tunneling term can be rewritten as
\be
\label{tunnel}
S_{\rm t}= t \int d^3x \ \left[ U^\dagger
\psi^\dagger_\uparrow \psi_\downarrow + 
 U \psi^\dagger_\downarrow \psi_\uparrow \right] \ . 
\ee
The tunneling of the electrons from one layer to another creates
a naked flux in the gauge field $a_-$, in other words, a monopole. 
If there were no monopole in Eq.\rf{tunnel}, we would have 
concluded that the tunneling just leads to the level splitting 
between the symmetric and antisymmetric sectors and 
proceeded from there. 
Here we need to take into account the monopole 
terms more carefully. 

In order to make a progress, we expand the partition 
function 
\be
\label{partition}
Z=\int {\cal D} a_+ {\cal D} a_- {\cal D} [\psi^\dagger_{+},\psi_{+}]
{\cal D} [\psi^\dagger_{-},\psi_{-}] \ e^{ - S - S_{\rm t} } \ ,
\ee
in powers of $t$ and evaluate the result of the expansion term 
by term and perform the $a_-$ integral in the presence
of monopoles.

In the expansion, each power of $U^\dagger$ and $U$ 
corresponds to creating a monopole and an antimonopole 
in the gauge field $a_-$ respectively. 
Let us show how the monopole contributions can be
evaluated following the analysis of \ct{KW}.
Consider, for example, the following correlation function which 
appears in the expansion of \rf{partition} to the second order in $t$, 
\be
\label{expa}
\VEV{U^\dagger(x) U(y) \ \psi^\dagger_\uparrow (x) \psi_\downarrow(x) 
\psi^\dagger_\downarrow(y)  \psi_\uparrow(y) } \ .
\ee
This corresponds to creating a monopole at $x$ and an antimonopole
at $y$ in addition to composite fermion creation and annihilation
processes. Since $U^{\dagger}$ generates a non-conserved
current coupled to the gauge field, naive integration over
$a_-$ becomes difficult. The situation can be improved
if we introduce the `monopole' current  
$j_\mu$ which is nonzero along the path $x(s)$ 
connecting the points $x$ and $y$ and has the following form,
\be
j_\mu(x) =\dot x_\mu(s) \delta(x-x(s)) \ .
\ee
Here two end points $x$ and $y$ corresponds to the 
positions of the monopole and antimonopole.
Now we can rewrite the correlation function \rf{expa} in the 
following fashion,
\bea
\label{smart}
\int {\cal D} a_{-} U^\dagger(x) U(y) e^{i \int d^3 x j_\mu a_{\mu -}}
 e^{ - S_{\rm eff}(a_-) } \br \times \left[ \VEV{
\psi^\dagger_\uparrow (x) \psi_\downarrow(x) 
\psi^\dagger_\downarrow(y)  \psi_\uparrow(y)} 
e^{-i\int d^3 x j_\mu a_{\mu -}} \right] \ , 
\eea
where 
\be
S_{\rm eff}(a_-) = - \ln \int 
{\cal D} [\psi^\dagger_+,\psi_+]
{\cal D} [\psi^\dagger_-,\psi_-] \ e^{-S} \ ,
\ee
and
\bea
\label{nonga}
\VEV{
\psi^\dagger_\uparrow (x) \psi_\downarrow(x) 
\psi^\dagger_\downarrow(y)  \psi_\uparrow(y)} = \br 
{\int {\cal D} [\psi^\dagger_+,\psi_+]
{\cal D} [\psi^\dagger_-,\psi_-] \ 
\psi^\dagger_\uparrow (x) \psi_\downarrow(x) 
\psi^\dagger_\downarrow(y)  \psi_\uparrow(y)
e^{-S} \over 
e^{-S_{\rm eff}(a_-)}} \ .
\eea
Thus the correlation function of Eq.(\ref{expa}) 
can be written as a product of two gauge invariant
quantities. In what follows we assume that the 
dependence of the term in the square braket of \rf{smart} 
on $a_-$ is weak and can be neglected. This may be a 
reasonable approximation for fermions which have typically
straight line paths. Now we have to evaluate the other 
factor with $U$, $U^\dagger$, and the current $j_\mu$.  

To do the $a_-$ integral in \rf{smart} we need to know $S_{\rm eff}$. 
One of the terms in this effective action is obviously the Chern-Simons
term, while the other term is generated dynamically and related
to the density and current response functions of the system with
respect to $a_-$ gauge field. 
The induced term is known in the RPA approximation.
%
Now the effective action in the Coulomb gauge can be written as
\be
\label{effective}
S_{\rm eff}(a)= {1 \over 2} \sum_{{\bf q},\omega} \sum_{\mu \nu}
a^*_\mu ({\bf q,\omega}) D^{-1}_{\mu \nu} ({\bf q},\omega)
a_{\nu} ({\bf q,\omega}) \ , 
\ee
and
\begin{eqnarray}
D^{-1}_{\mu \nu} = \left (
\begin{array}{cc}
-{m^* \over 2\pi}\left (1+i{\omega \over v_Fq} \right), 
{iq \over  2 \pi \phi_0} \\
-{iq \over  2 \pi \phi_0}, 
-i{2 n_e \over k_F}{\omega \over q} + \chi q^2
\end{array}
\right ) \ ,  
\end{eqnarray}
where 
$\chi=1/(24 \pi m^*) + 
(1 + e^2m^*d/\epsilon)/(2\pi m^* \phi^2_0)$.
Here $d$ is the layer spacing and $\epsilon$ is
the static dielectric constant of the system.
Solving the equation of motion
\be
\label{eq}
\sum_{\nu} D^{-1}_{\mu \nu} ({\bf q,\omega})
a_{\nu} ({\bf q,\omega}) = j_\mu ({\bf q,\omega}) \ ,
\ee
and substituting the solution into the effective action \rf{effective}
the first factor, $U^{\dagger}(x)U(y)$, 
in Eq.(\ref{expa}) becomes $e^{-S_{M{\bar M}}}$
where $S_{M{\bar M}}$ is the action for the monopole and 
antimonopole sitting at $x$ and $y$ respectively.
$S_{M{\bar M}}$ depends on the distance in space-time 
between the monopole and antimonopole. For example, one
can show that 
\be
\label{inst}
S_{M{\bar M}} (|x_0-y_0|,{\bf x}={\bf y}) 
\propto |x_0-y_0|^{1/3} \ , 
\ee
where $|x_0-y_0|$ is the distance in time.
We will see that the detailed from of
$S_{M{\bar M}}$ is not very important
as far as it is finite. 

Repeating these manipulations for other terms of the 
perturbative expansion of \rf{partition} in powers of $t$,
we can make the following observations. First, the terms
in the odd powers of $t$ in this expansion will involve
isolated monopoles or antimonopoles whose action diverges.
Thus these terms are completely suppressed. 
%
Second, among the terms with even powers of $t$, 
only the terms which have equal number of monopoles 
and antimonopoles survive. For each pair of the monopole
and antimonopole, $e^{-S_{M {\bar M}}}$ factor must
be augmented. 
It can be shown that all these terms can be resummed 
to give an effective four-fermion interaction 
\begin{equation}
\tilde S_{\rm t} = t^2 \int d^3 x \ d^3 y \ 
\psi^\dagger_\uparrow (x) \psi_\downarrow (x)
\psi^{\dagger}_\downarrow (y) \psi_\uparrow (y) 
e^{-S_{M {\bar M}}} \ .
\end{equation}
Using the Hubbard-Stratonovic field $Q(x)$,
we can rewrite the above interaction term as
\begin{eqnarray}
\label{tunew}
\tilde S_{\rm t} &=& t \int d^3 x \ [ Q(x) 
\psi^\dagger_\uparrow \psi_\downarrow + Q^* (x) 
\psi^\dagger_\downarrow \psi_\uparrow ] \cr 
&&+ \ {1 \over 2} \int d^3 x \ d^3 x' \ Q(x)
\hat K(x-x') Q^*(x') \ ,
\end{eqnarray}
where $\hat K(x)$ can be determined from
$e^{-S_{M {\bar M}}}$.
Note that finite $Q$ will provide a dynamically
generated {\it effective} composite fermion tunneling 
term and it will lead to the transition to the pseudospin
polarized state. 
Without loss of generality, we can take $Q$ to be real.
One can show that any finite imaginary component corresponds to
choosing an effective magnetic field along a different
direction in pseudospin space. 

Now we face the theory defined by the sum of the mean field
action and the new interaction term \rf{tunew}. 
\begin{eqnarray}
S &=&
\int d^3 x \ 
\left [ \psi^{\dagger}_{+} \left ( \partial_{\tau} - 
{\nabla^2 \over 2m^*} - \mu + t Q(x) \right ) 
\psi_{+} \right. \cr
&&\left. \psi^{\dagger}_{-} \left ( \partial_{\tau} - 
{\nabla^2 \over 2m^*} - \mu - t Q(x) \right ) 
\psi_{-} \right ] \cr
&&+ \ {1 \over 2} 
\int d^3 x \ d^3 x' \ Q(x) \hat K(x-x') Q(x') \ .
\end{eqnarray}
We are going to apply the saddle point approximation to the
action to determine $Q$ field. The saddle point action is
given by
\begin{eqnarray}
S_{\rm SP} &=& - 
\sum_{{\bf k},\omega} \left [ 
\psi^{\dagger}_{{\bf k}+} (i\omega - \xi_{\bf k} - tQ) 
\psi_{{\bf k}+} \right. \cr
&&+ \ \left.
\psi^{\dagger}_{{\bf k}-} (i\omega - \xi_{\bf k} + tQ) 
\psi_{{\bf k}-} \right ] \cr
&&+ \ {1 \over 2} \ Q^2
\int d^3 x \ d^3 x' \ \hat K(x-x') \ .
\end{eqnarray}
where $\xi_{\bf k} = k^2/2m^* - \mu$.
After integrating out fermions and minimizing the resulting
action with respect to $Q$, 
we find the following self-consistent equation (we set the
volume of the system to be unity).
\begin{equation}
\label{saddle2}
t \sum_{\bf k} \left [ f(\xi_{{\bf k}-}) - f(\xi_{{\bf k}+}) \right ] 
= K Q \ , 
\label{self}
\end{equation}
where $\xi_{\bf k\pm} = \xi_{\bf k} \pm t Q$, 
$K=\int d^3 x \hat K(x)$, and $f(x)=1/(e^{x/T}+1)$ is
the Fermi distribution function.

At the zero temperature, if $Qt < \mu$, the self-consistent 
equation becomes
\be
t N(0) (2tQ) = KQ \ ,
\ee
where $N(0)=m^*/2\pi$ is the density of states in 
two dimensions. Thus there is only the trivial 
solution $Q = 0$ in this case. This means that 
partially pseudospin polarized states are not possible.
However, if $Qt > \mu$, only the antisymmetric sector 
will be occupied, thus it becomes
\be
t n_{\rm total} = KQ \ ,
\ee
where $n_{\rm total}$ is the total density of 
electrons. This implies that $Qt > \mu$ can be satisfied
as far as $t > t_0 = \sqrt{K/2N(0)}$.
Therefore, the fully pseudospin polarized state 
is possible if the tunneling strength is larger
than the threshold value $t_0 = \sqrt{K/2N(0)}$. 
Since partially pseudospin polarized states are
not possible, we conclude that the transition to 
the fully polarized phase is of first order.
One can also show that finite temperature corrections
are small as far as $T < \mu$ and the above conclusion
is still valid. 

As mentioned in the introduction, some pairing instability
may derive the pseudospin unpolarized and polarized 
Fermi-liquid-like states to the (331) and MR states 
respectively when the interlayer interaction becomes
stronger. 
In view of the suggestion that the transition
from the (331) state to the MR state is continuous,
it will be interesting to understand how the first
order transition found here is replaced by the 
continuous transition when the paired quantum Hall 
states are formed.  

In summary, we studied the effect of tunneling in the double
layer Fermi-liquid-like states. There exists the first order
transition from two separate Fermi-liquid-like states with
$\nu=1/4$ to a single Fermi-liquid-like state with $\nu=1/2$
when the tunneling amplitude becomes sufficiently large. 

This work was supported by the NSF grant PHY-94-07194 (V.G.)
and Sloan Foundation Fellowship (Y.B.K.).
We are also grateful to Aspen center for physics where the
present work was initiated.
We would like to thank Sasha Abanov, Matthew Fisher, 
Jason Ho, Ed Rezayi,
Xiao-Gang Wen, and Tony Zee for helpful discussions.

\end{multicols}

\end{document}